\documentclass[a4paper,12pt]{article}

\usepackage{amsmath}
\usepackage{amssymb}
\usepackage{mathrsfs}
\usepackage[latin1]{inputenc}
\usepackage{graphicx,graphics}
\usepackage{epsf}
\usepackage{a4wide,latexsym,cite,verbatim}
\usepackage[usenames]{color}

\newtheorem{theorem}{Theorem}
\newtheorem{corollary}{Corollary}
\newtheorem{lemma}{Lemma}
\newtheorem{proposition}{Proposition}

\newtheorem{claim}{Claim}

\def\qed{\hspace{0.4cm}\rule{.2cm}{.2cm}\vspace{\baselineskip}}

\newenvironment{proof}[1][Proof. ]
    {\noindent\textbf{#1}}
    {\hspace{0.4cm}\rule{.2cm}{.2cm}\vspace{\baselineskip}}

\def\F{ {\mathcal F} }

\def\C{ {\mathcal C} }

\def\S{ {\mathcal S} }

\begin{document}

\begin{center}
{\bf \Large Clique cycle-transversals in distance-hereditary graphs}\\[5mm]
{\bf Andreas Brandst\"adt}$^1$,\\
{\bf Simone Esposito}$^2$,\\
{\bf Loana Tito Nogueira}$^2$,\\
and\\
{\bf Fábio Protti}$^2$\\[5mm]
$^1$
Institut für Informatik,
Universität Rostock,
Germany,
e-mail:{\tt ab@informatik.uni-rostock.de}\\[3mm]
$^2$
Instituto de Computa\c c\~ao,
Universidade Federal Fluminense,
Niter\'oi, RJ,
Brazil,\\
email: {\tt \{loana,fabio\}@ic.uff.br, simone.esposito@light.com.br}\\[3mm]
\end{center}

\bigskip

\noindent {\bf Abstract.} A {\em cycle-transversal} of a graph $G$ is a subset $T \subseteq V(G)$ such that
$T \cap V(C) \neq \emptyset$ for every cycle $C$ of $G$. A {\em clique cycle-transversal}, or {\em cct} for short, is a cycle-transversal which is a clique. Recognizing graphs which admit a cct can be done in polynomial time; however, no structural characterization of such graphs is known. We characterize distance-hereditary graphs admitting a cct in terms of forbidden induced subgraphs. This extends similar results for chordal graphs and cographs.

\section{Introduction}

A {\em cycle-transversal} of a graph $G$ is a subset $T \subseteq V(G)$ such that $T \cap V(C) \neq \emptyset$ for every cycle $C$ of $G$. When $T$ is a clique, we say that $T$ is a {\em clique cycle-transversal} or simply {\em cct}. A graph admits a cct if and only if it can be partitioned into a complete subgraph and a forest; by this reason such a graph is also called a $(\C,\F)$-{\it graph} in \cite{BraBriKleNogPro2011}.

Finding a minimum cycle-transversal in a graph is NP-hard due to a general result in~\cite{yanna}, which says that the problem of finding the minimum number of vertices of a graph $G$ whose deletion results in a subgraph satisfying a hereditary property $\pi$ on induced subgraphs is NP-hard. This result implies the NP-hardness of other problems involving cycle-transversals, for instance the problem of finding a minimum odd cycle-transversal (which is equivalent to finding a maximum induced bipartite subgraph), or the problem of finding a minimum triangle-transversal (which is equivalent to finding a maximum induced triangle-free subgraph). Odd cycle-transversals are interesting due to their connections to perfect graph theory; in~\cite{reed}, an $O(mn)$ algorithm is developed to find odd cycle-transversals with bounded size. In~\cite{dmtcs}, the authors study the problem of finding $C_k$-transversals, for a fixed integer $k$, in graphs with bounded degree; among other results, they describe a polynomial-time algorithm for finding minimum $C_4$-transversals in graphs with maximum degree three.

Graphs admitting a cct can be recognized in polynomial time, as follows. Note first that $(\C,\F)$-graphs form a subclass of $(2,1)$-graphs (graphs whose vertex set can be partitioned into two stable sets and one clique). The strategy for recognizing a $(\C,\F)$-graph $G$ initially checks whether $G$ is a $(2,1)$-graph, which can be done in polynomial time (see~\cite{B96}). If so, then test, for each candidate clique $Q$ of a $(2,1)$-partition of $G$, if $G-Q$ is acyclic (which can be done in linear time). If the test fails for all cliques $Q$, then $G$ is not a $(\C,\F)$-graph, otherwise $G$ is a $(\C,\F)$-graph. To conclude the argument, we claim that the number of candidate cliques $Q$ is polynomial. Since $G$ is a $(2,1)$-graph, let $(B,Q)$ be a $(2,1)$-partition of $V(G)$ where $B$ induces a bipartite subgraph and $Q$ is a clique. Let $(B',Q')$ be another $(2,1)$-partition of $V(G)$. Then $|Q'\setminus Q| \leq 2$ and $|Q\setminus Q'| \leq 2$, otherwise $G[B]$ or $G[B']$ would contain a triangle, which is impossible. Therefore, we can generate in polynomial time all the other candidate cliques $Q'$ from $Q$. This is the same argument used to count sparse-dense partitions (for more details see \cite{FHKM99}).
Although recognizing graphs admitting a cct can be done in polynomial time, no structural characterization of such graphs is known, even for perfect graphs.

A similar sparse-dense partition argument can be employed to show that an interesting superclass of $(\C,\F)$-graphs, namely graphs admitting a clique triangle-transversal, can also be recognized in polynomial time. Such graphs are also known in the literature as $(1,2)${\em -split graphs}. A characterization of this
class is given in~\cite{ZZ01}, where it has been proved that there are $350$ minimal forbidden induced subgraphs for $(1,2)$-split graphs. When $G$ is a perfect graph, being a $(1,2)$-split graph is equivalent to
being a $(2,1)$-graph: observe that a perfect graph $G$ contains a clique triangle-transversal if and only if $G$ contains a clique that intersects all of its odd cycles. In~\cite{BNK05} and~\cite{Loana}, respectively, characterizations by forbidden induced subgraphs of cographs and chordal graphs which are $(1,2)$-split
graphs are presented.

\if 10

Regarding $(\S,\F)$-graphs, in the literature they are also called {\em near-bipartite graphs}.
In~\cite{YY06}, the authors prove that recognizing $(\S,\F)$-graphs is NP-complete
even for graphs of maximum degree four or for graphs of diameter four.
Moreover, they presented simple characterizations for $(\S,\F)$-graphs
of maximum degree at most three, and for $(\S,\F)$-graphs of diameter two.
(As far as the authors know, the complexity of recognizing $(\S,\F)$-graphs of diameter three is still open.)
Independently, the NP-completeness of recognizing graphs partitionable
into an edgeless graph and a tree has been proved in~\cite{BLS89}.

A natural superclass of $(\S,\F)$-graphs consists of the graphs which
admit a stable triangle-transversal. Recognizing such graphs
is NP-complete, as an easy consequence of a result in~\cite{Farrugia}.
There are infinitely many vertex-minimal graphs which do not admit
an stable triangle-transversal, for instance the powers of cycles $C^2_n$, for
$n=5+3k$ and $n=7+3k$, $k\geq0$. However, the problem of deciding whether a perfect graph admits a stable
triangle-transversal corresponds trivially to the
$3$-colorability problem: note that a perfect graph admits such a transversal
if and only if there exists a stable set whose removal leaves a
graph containing no $C_{2k+1}$ for $k \geq 1$, i.e., a bipartite
graph. In other words, a perfect graph $G$ admits a stable triangle-transversal
if and only if $G$ does not contain $K_4$. For a chordal graph $G$, $S$ is a stable
triangle-triangle transversal if and only if $S$ is an sct.

In the section of conclusions, we describe infinite families of
vertex-minimal distance-hereditary graphs which do not admit an sct, as an
evidence of the difficulty in recognizing distance hereditary graphs which are
$(\S,\F)$-graphs.

\fi

Deciding whether a distance-hereditary graph admits a cct can be done in linear-time using the clique-width approach, since the existence of a cct can be represented by a Monadic Second Order Logic (MSOL) formula using only predicates over vertex sets~\cite{courcelle,rao}. However, no structural characterization for distance-hereditary graphs admitting a cct was known. In order to fill this gap, in this note we describe a characterization of distance-hereditary graphs with cct in terms of forbidden induced subgraphs.

\section{Background}

In this work, all graphs are finite, simple and undirected. Given
a graph $G=(V(G),E(G))$, we denote by $\overline{G}$ the
complement of $G$. For $V' \subseteq V(G)$, $G[V']$ denotes the
subgraph of $G$ induced by $V'$. Let $X=(V_X, E_X)$ and $Y=(V_Y,
E_Y)$ be two graphs such that $V_X \cap V_Y = \emptyset$. The
operations ``+" and ``$\cup$" are defined as follows: the {\em
disjoint union} $X \cup Y$, sometimes referred simply as {\em
graph union}, is the graph with vertex set $V_X \cup V_Y$ and edge
set $E_X \cup E_Y$; \ the {\em join} $X+Y$ is the graph with
vertex set $V_X \cup V_Y$ and edge set $E_X \cup E_Y \cup \{xy
\mid x \in V_X, y \in V_Y\}$.

Let $N(x)=\{y \mid y \neq x$ and $xy \in E\}$ denote the open neighborhood of $x$ and let $N[x]=\{x\} \cup N(x)$ denote the closed neighborhood of $x$. A {\em cut-vertex} is a vertex $x$ such that $G[V \setminus \{x\}$ has more connected components than $G$. A {\em block} (or {\em $2$-connected component}) of $G$ is a maximal induced subgraph of $G$ having no cut-vertex. A block is {\em nontrivial} if it contains a cycle; otherwise it is {\em trivial}.

For a set ${\cal F}$ of graphs, $G$ is {\em ${\cal F}$-free} if no induced subgraph of $G$ is in ${\cal F}$.

Vertices $x$ and $y$ are {\em true twins} ({\em false twins}, respectively) in $G$ if $N[x]=N[y]$ ($N(x)=N(y)$, respectively).

{\em Adding a true twin} ({\em false twin, pendant vertex}, respectively) $y$ to vertex $x$ in graph $G$ means that for $G$ and $y \notin V(G)$, a new graph $G'$ is constructed with $V(G')=V(G) \cup \{y\}$ and $E(G')=E(G) \cup \{xy\} \cup \{uy \mid u \in N(x)\}$ ($E(G')=E(G) \cup \{uy \mid u \in N(x)\}$, $E(G')=E(G) \cup \{xy\}$, respectively).

The {\it complete} (resp. {\it edgeless}) graph with $n$ vertices is denoted by $K_n$ (respectively
$I_n$). The graphs $K_1$ and $K_3$ are called {\it trivial graph} and {\it triangle},
respectively. The {\em chordless cycle} ({\em chordless path}, respectively) with $n$ vertices is denoted by $C_n$ ($P_n$, respectively). The graph $C_n$ ($\overline{C_n}$, respectively) for $n\geq 5$ is a {\it hole} ({\it anti-hole}, respectively).

The {\it house} is the graph with vertices $a,b,c,d,e$ and edges $ab,bc,cd,ad,ae,be$.
The {\it gem} is the graph with vertices $a,b,c,d,e$ and edges $ab,bc,cd,ae,be,ce,de$.
The {\it domino} is the graph with vertices $a,b,c,d,e,h$ and edges $ab,bc,cd,ad,be,eh,ch$.

\medskip

\begin{figure}[hdg]
\centering
\includegraphics[width=10cm]{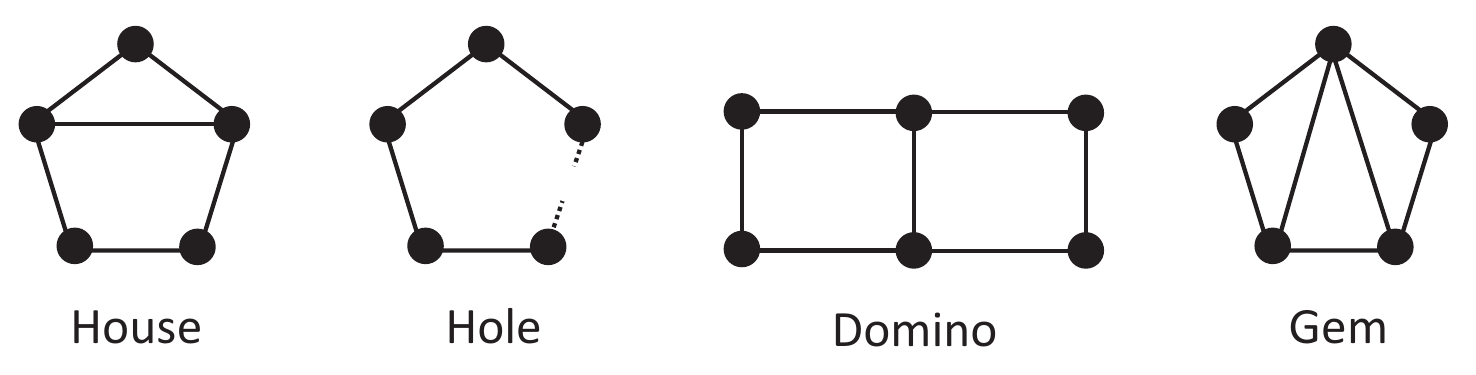}
%\vspace{-1cm}
\caption{House, hole, domino, and gem.}\label{householedominogem}
\end{figure}

If $H$ is an induced subgraph of $G$ then we say that $G$ {\em contains} $H$,
otherwise $G$ is $H$-{\it free}. A {\it clique} (resp. {\it stable} or
{\it independent set}) is a subset of vertices inducing a complete
(resp. edgeless) subgraph. A {\it universal vertex} is a vertex
adjacent to all the other vertices of the graph. A {\it split graph}
is a graph whose vertex set can be partitioned into a stable set and a clique.
It is well known that $G$ is a split graph if an only if $G$ is $(2K_2,C_4,C_5)$-free.

A {\it star} is a graph whose vertex set can be partitioned into a stable set and a universal vertex.
A {\it bipartite graph} is a graph whose vertex set can be partitioned into two stable sets.
A {\it cograph} is a graph containing no $P_4$. A {\it chordal graph} is a graph
containing no $C_k$, for $k \geq 4$. A {\it distance-hereditary graph}
is a graph in which the distances in any connected induced subgraph
are the same as they are in the original graph. A {\it threshold graph}
is a graph that can be constructed from a one-vertex graph by repeated applications of the following two operations:
(a) addition of a single isolated vertex to the graph;
(b) addition of a single universal vertex to the graph. It is well known that $G$ is a threshold graph if an only if $G$ is $(2K_2,C_4,P_4)$-free.
%A {\it perfect graph} is a graph containing no odd holes/anti-holes.

If $T \cap V(C) \neq \emptyset$ for cycle $C$, we say that $T$ {\em covers} $C$.

\section{The forbidden subgraph characterization}

%\subsection{Some useful properties of distance-hereditary graphs}

The following well-known characterization of distance-hereditary graphs, also called {\em HHDG-free graphs}, will be fundamental for our result:

\begin{theorem}{\em \cite{BanMul1986}}\label{maindhg}
The following are equivalent for any graph $G$:
\begin{itemize}
\item[$(i)$] $G$ is a distance-hereditary graph.
\item[$(ii)$] $G$ can be generated from a single vertex by repeatedly adding a pendant vertex, a false twin, or a true twin, respectively.
\item[$(iii)$] $G$ is $($house, hole, domino, gem$)$-free. (See Figure \ref{householedominogem}.)
\end{itemize}
\end{theorem}

Let $G=(V,E)$ be a graph, and for a vertex $x \in V$, let $N^k(x)=\{y \in V \mid dist_G(x,y)=k\}$ for $k \ge 0$ denote the distance levels in $G$ with respect to $x$. For $k=2$, let $R= V \setminus (N[x] \cup N^2(x))$. The following are useful properties of distance-hereditary graphs:

\begin{proposition}\label{sameneighbors}
Let $G$ be distance hereditary and $u,v \in N^2(x)$.
\begin{itemize}
\item[$(i)$] If $uv\in E(G)$ or $uv\not\in E(G)$ but connected by a path in $N^2(x) \cup R$ then $N(u)\cap N(x) = N(v)\cap N(x)$.
\item[$(ii)$] If $N(x)$ is a stable set then for $u,v \in N^2(x)$ with $uv \notin E$, $N(u) \cap N(x)$ and $N(v) \cap N(x)$ do not overlap.
\end{itemize}
\end{proposition}

{\bf Proof.} $(i)$: Since $G$ is (house,hole)-free, $u,v$ cannot have incomparable neighborhoods in $N(x)$. Moreover, since $G$ is (house, hole, domino, gem)-free, the neighborhoods of $u$ and $v$ in $N(x)$ cannot properly contain one another, which shows Proposition~\ref{sameneighbors}.

$(ii)$: Let us suppose, by contradiction, that $u$ and $v$ overlap. In this case, let $a, b, c$ be vertices in $N(x)$ such that $au \in E(G)$, $av \notin E(G)$, $bu, bv \in E(G)$, $cu \notin E(G)$ and $cv \in E(G)$. Then $G[x, a, b,c, u, v]$ induces a domino, which is a contradiction. \qed

\begin{figure}[tb]
\includegraphics[width=16cm]{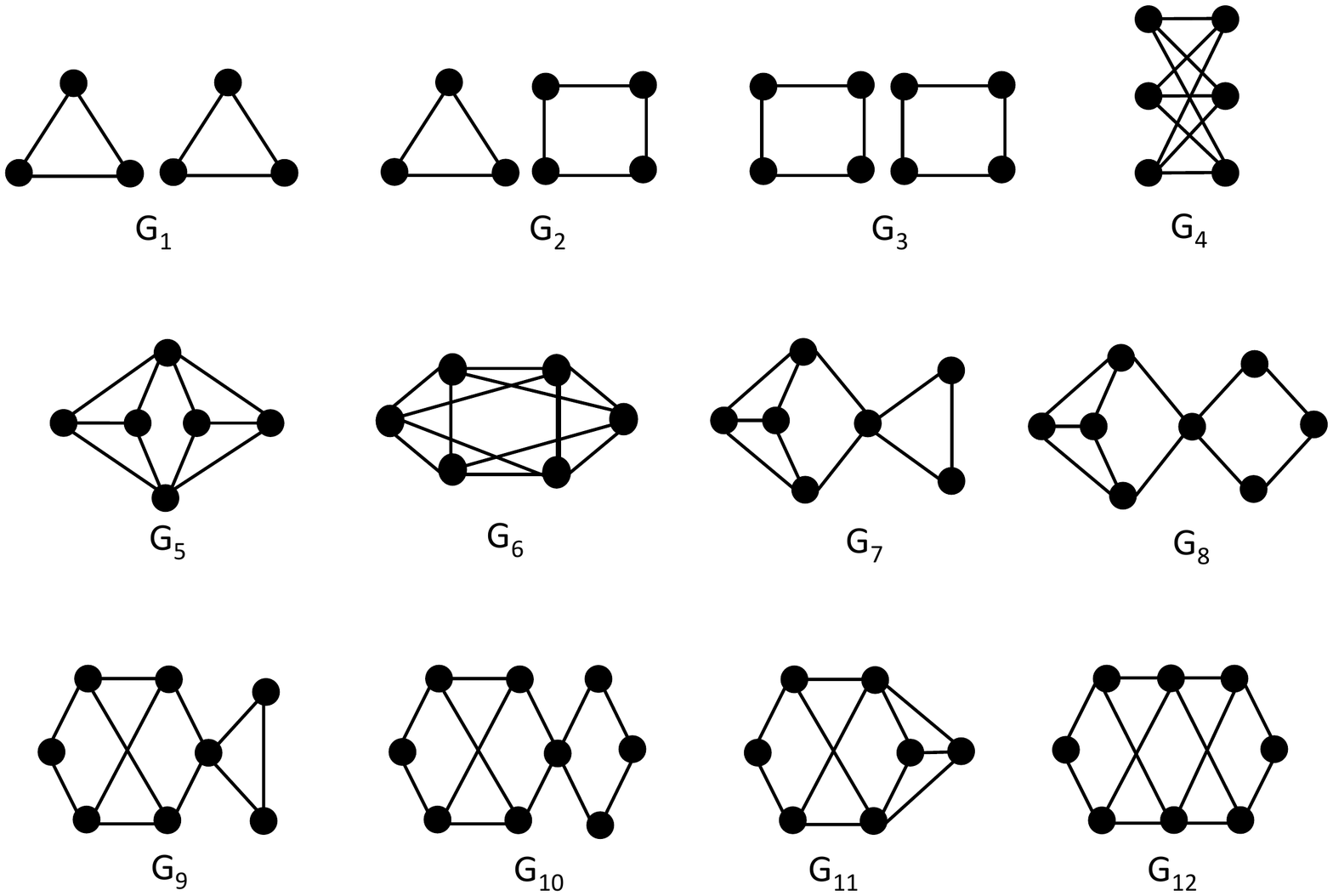}
\vspace{-1cm}
\caption{Forbidden subgraphs for distance-hereditary graphs with cct.}\label{cctdhg}
\end{figure}

\if 10

\begin{lemma}\label{blocks}
Let $G'=(V',E')$ be a $(G_1,\ldots,G_{12})$-free distance-hereditary graph resulting from $G=(V,E)$ by adding a false twin $y \notin V$ to vertex $x \in V$.
If $x$ is contained in a nontrivial block of $G$ and $x$ is a cut-vertex of $G$ then all other blocks of $G$ are trivial.
\end{lemma}

{\bf Proof.}
Since $G'$ is $(G_1,G_2,G_3$-free, the distance between nontrivial blocks is at most one. If $x$ is in more than one nontrivial blocks then, since $G'$ is hole-free, $x$ is in more than one $C_3$ or $C_4$ which share only $x$ as common vertex but then, together with $y$, there is $G_5$ or $G_{11}$ or $G_{12}$ in $G'$ which is impossible. If $x$ is in a nontrivial block $B$ of $G$ and there is a nontrivial block $B'$ of $G$ in distance one from $x$ then, together with $y$, there is $G_5$ or $G_7$ or $G_8$ or $G_9$, $G_{10}$ or $G_{11}$  in $G'$ which is again impossible. This shows Lemma \ref{blocks}.
\qed

\fi

\begin{theorem}\label{maintheocctdhg}
Let $G$ be a distance-hereditary graph. Then $G$ admits a clique cycle transversal if and only if $G$ is $(G_1,\ldots,G_{12})$-free.
\end{theorem}

\proof It is easy to see that $G_1,\ldots,G_{12}$ from Figure \ref{cctdhg} have no cct. For the converse direction, let $G'$ be a distance-hereditary $(G_1,\ldots,G_{12})$-free graph. By Theorem \ref{maindhg}, $G'$ results, starting with a single vertex, by repeatedly applying one of the three operations in Theorem \ref{maindhg} $(ii)$. Since adding a pendant vertex $y$ to a vertex $x$ in $G$ does not create cycles with $y$, we can restrict ourselves to the following two cases: $G'$ results from $G$ by either adding a true twin or a false twin $y$ to vertex $x$ in $G$, and in both cases, we have to show that $G'$ has a cct.

We can inductively assume that $G$ has a cct $Q$. The vertex set $V(G)$ can be partitioned into $\{x\} \cup N(x) \cup N^2(x) \cup R$.
Let $Q_1=Q \cap N(x)$, $Q_2=Q \cap N^2(x)$ and $N_1(x)=N(x)\setminus Q_1$.

\subsection{Case 1: $y$ is a true twin to $x$.}

Let $G'$ result from $G$ by adding a true twin $y$ to $x$ in $G$.
In this case, the possible cycles with $y$ in $G'$ are triangles $xya$ for $a \in N(x)$, triangles $yab$ for $a,b \in N(x), ab \in E(G)$, and $C_4$'s $yabc$ for $a,b \in N(x)$, $ab \notin E(G)$, $c \in N^2(x)$. If $x \in Q$ or, more generally, $Q \subseteq N[x]$, then $Q \cup \{y\}$ is a cct of $G'$. Thus we have to consider the case $x \notin Q$. Since for a triangle $yab$ also $xab$ is a triangle which is covered by $Q$, the triangle $yab$ is covered by $Q$, and similarly for the $C_4$ $yabc$ where $xabc$ is a $C_4$ in $G$ covered by $Q$. Thus, we only have to deal with triangles $xya$.

\begin{claim}\label{N(x)split}
If $x\not\in Q$ then $G[N(x)]$ is a split graph with partition $(N_1(x), Q_1)$.
\end{claim}

{\em Proof of Claim $\ref{N(x)split}$.} For each edge $ab \in G[N(x)]$, $xab$ is a triangle. Hence, $a$ or $b$ is in $Q$ and $N_1(x)$ is a stable set. Since $Q_1$ is a clique, the claim follows. \hfill $\diamond$

Since $G'$ is $(G_1,G_2,G_3)$-free, $R$ induces a cycle-free subgraph in $G$.

\medskip

%\subsubsection{Case 1.1: $G[N^2(x)\cup R]$ is cycle-free.}
{\bf Case 1.1: $G[N^2(x)\cup R]$ is cycle-free.}

\begin{claim}\label{ttN2(x)URcyclefree}
If
%$y$ is a true twin to $x$,
$x\not\in Q$
%and $G[N^2(x)\cup R]$ is cycle-free
then $G'$ has a cct.
\end{claim}

{\em Proof of Claim $\ref{ttN2(x)URcyclefree}$.} Since $Q$ is a clique and $G[N^2(x)\cup R]$ is cycle-free, $Q_2$ contains at most two vertices. If $Q_2=\emptyset$ then $Q \cup \{y\}$ is a cct of $G'$. If a vertex $u \in Q_2$ has no neighbors in $N_1(x)$ then every cycle containing $u$ also contains a vertex of $Q_1$, i.e., $Q\setminus\{u\}$ is still a cct of $G$. Thus, assume without loss of generality that every vertex in $Q_2$ has a neighbor in $N_1(x)$.

If $Q_2=\{u,v\}$, the neighborhood of $Q_2$ in $N_1(x)$ cannot contain two vertices $a$ and $b$, otherwise by Claim~\ref{N(x)split} and Proposition~\ref{sameneighbors} vertices $x,y,a,b,u,v$ induce $G_5$. Hence $u$ and $v$ have precisely one neighbor $a \in N_1(x)$ which must be adjacent to all vertices of $Q_1$, otherwise if $a$ misses a vertex $b \in Q_1$ then vertices $x,y,a,b,u,v$ induce $G_5$. Therefore $(Q\setminus\{u,v\})\cup\{a,y\}$ is a cct of $G'$.

If $Q_2=\{u\}$, we consider two subcases:

$(i)$ Vertex $u$ has a neighbor $a \in N_1(x)$ which misses some vertex $b \in Q_1$. Then every cycle $C$ in $G$ containing $a,u$ but no vertex of $Q_1$ must also contain $x$. This is shown as follows: If $C$ does not contain $x$ then, by Proposition~\ref{sameneighbors}, $C$ is either a triangle $auv$ with $v\in N^2(x)$ or a $C_4$ $aucv$ with $c\in N_1(x), v\in N^2(x)$. In the former case by using Proposition~\ref{sameneighbors}, vertices $x,y,a,b,u,v$ induce $G_5$.
The latter case cannot occur since the existence of cycle $aucv$ implies the existence of cycle $axcv$ in $G$, not covered by $Q$. This implies that $(Q\setminus\{u\})\cup\{x,y\}$ is a cct of $G'$.

$(ii)$ Every neighbor $a \in N_1(x)$ of $u$ sees all vertices in $Q_1$. Then $Q\cup\{a\}$ is a cct of $G$ for some $a \in N_1(x)$ and, since by Claim~\ref{N(x)split} $N_1(x)$ is a stable set, every other neighbor $a' \in N_1(x)$ of $u$ misses some vertex in $Q_1\cup\{a\}$. By applying a similar argument as in $(i)$, every cycle $C$ in $G$ containing $a',u$ but no vertex of $Q_1\cup\{a\}$ must also contain $x$. We conclude that $(Q\setminus\{u\})\cup\{x,y,a\}$ is a cct of $G'$. This completes the proof of Claim~\ref{ttN2(x)URcyclefree}. \hfill $\diamond$

%\subsubsection{Case 1.2: $G[N^2(x)\cup R]$ is not cycle-free.}
\bigskip

{\bf Case 1.2: $G[N^2(x)\cup R]$ is not cycle-free.}

\begin{figure}[h]
\begin{center}
\includegraphics[width=13cm]{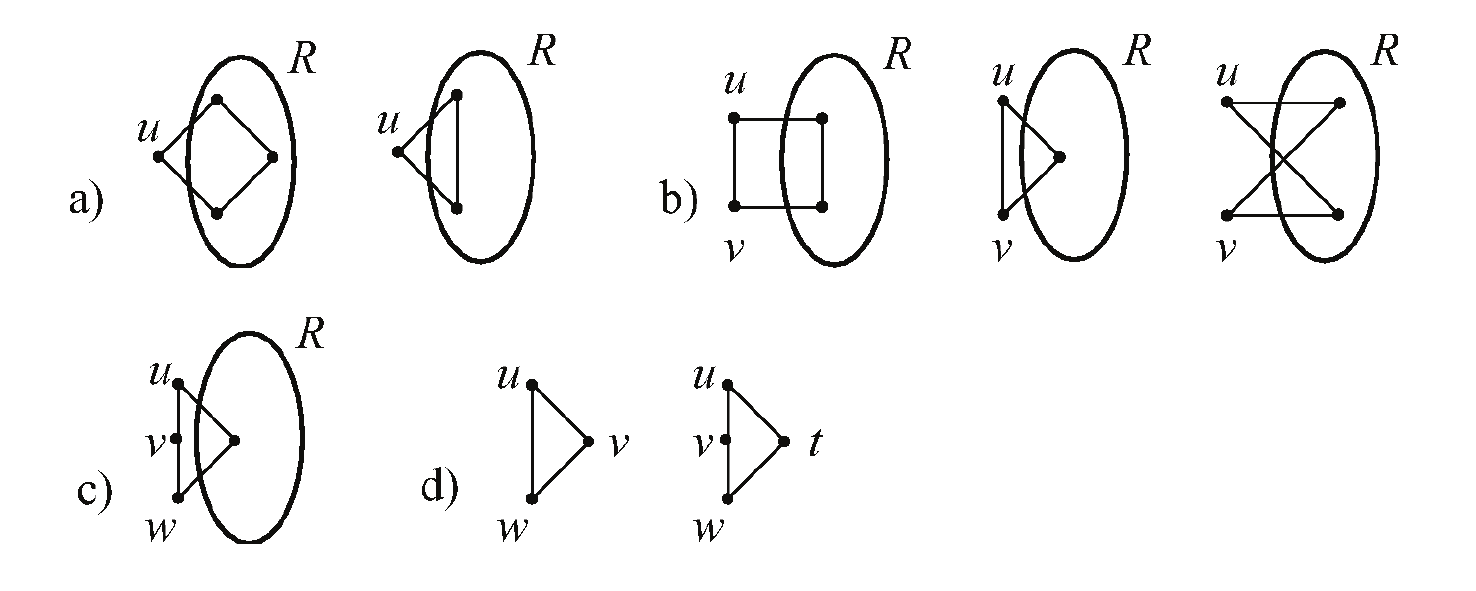}
\caption{Cycles in $G[N^2(x) \cup R]$.}\label{cycnxr}
\end{center}
\end{figure}

We now assume that $G[N^2(x)\cup R]$ contains a cycle $C$. This cycle can be one of the following types (see Figure \ref{cycnxr}):

\begin{itemize}
\item[($A_1$)] $C$ has exactly one vertex $u$ in $N^2(x)$, and $C$ is a $C_4$.
\item[($A_2$)] $C$ has exactly one vertex $u$ in $N^2(x)$, and $C$ is a $C_3$.
\item[($B_1$)] $C$ has exactly two vertices $u,v$ in $N^2(x)$, $uv \in E(G)$, and $C$ is a $C_4$.
\item[($B_2$)] $C$ has exactly two vertices $u,v$ in $N^2(x)$, $uv \in E(G)$, and $C$ is a $C_3$.
\item[($B_3$)] $C$ has exactly two vertices $u,v$ in $N^2(x)$, $uv \notin E(G)$, and $C$ is a $C_4$.
\item[($C_1$)] $C$ is a $C_4$ with exactly three vertices $u,v,w$ in $N^2(x)$ (which form a $P_3$ in $N^2(x)$).
\item[($D_1$)] $C$ is a $C_3$ in $N^2(x)$.
\item[($D_2$)] $C$ is a $C_4$ in $N^2(x)$.
\end{itemize}

\begin{claim}\label{ttN(X)clique}
%If $y$ is a true twin to $x$ and there is a cycle in $G[N^2(x)\cup R]$ then
$N(x)$ is a clique.
\end{claim}

{\em Proof of Claim $\ref{ttN(X)clique}$.}
Suppose to the contrary that there are $a,b \in N(x)$ with $ab \notin E(G)$. Since $G'$ is $(G_1,G_2)$-free, $a$ and $b$ must see each cycle in $G[N^2(x)\cup R]$. If $C$ is a cycle of type ($A_1$) or ($A_2$) in $G[N^2(x)\cup R]$, i.e., with exactly one vertex $u$ in $N^2(x)$ then $a$ and $b$ see $u$ and we obtain $G_7$ or $G_8$ - contradiction. If $C$ is of type ($B_3$) with $u,v, \in N^2(x)$, $uv\notin E(G)$, then both $a$ and $b$ have to see $C$, and if not both $a$ and $b$ see both $u$ and $v$ then there is either a hole or domino or $G_8$. Thus $a$ and $b$ see both $u$ and $v$, i.e., there is $G_{11}$ - contradiction.

We analyze the remaining cases by considering the following situation: If $a$ sees vertex $u$ and $b$ sees vertex $v\neq u$ in $N^2(x)$ such that there exists a path linking $u$ and $v$ in $N^2(x)$ then by Proposition~\ref{sameneighbors}, $a$ and $b$ see a common edge $u'v'$; but then $G'$ contains $G_5$ with $x,y,a,b,u',v'$ - contradiction. This shows Claim \ref{ttN(X)clique}. \hfill $\diamond$

Let $N^2_C(x)$ denote the set of all vertices in $N^2(x)$ which are contained in cycles of subgraph $G[N^2(x) \cup R]$. Since $G$ is $(G_1,G_2,G_3)$-free, there is only one connected component in $G[N^2_C(x) \cup R]$. In addition, if $a \in N(x)$ then there is a triangle $xya$, and $a$ must see every cycle in $G[N^2_C(x) \cup R]$.

\begin{claim}\label{joinN(x)Q'2}
%If $y$ is a true twin to $x$ and there is a cycle in $G[N^2(x)\cup R]$ then
Every vertex in $N(x)$ sees every vertex in $Q_2$.
\end{claim}

{\em Proof of Claim $\ref{joinN(x)Q'2}$.} Since $G'$ is $(G_1,G_2,G_3)$-free, any vertex $a \in N(x)$ sees at least one vertex $u$ in each cycle of $G[N^2(x) \cup R]$. Since by Claim~\ref{ttN(X)clique}, $N(x)$ is a clique, and by Proposition~\ref{sameneighbors}, all vertices in $Q_2$ have the same neighborhood in $N(x)$, and vertex $u$ sees $a$, all vertices in $Q_2$ see all vertices in $N(x)$ which shows Claim \ref{joinN(x)Q'2}. \hfill $\diamond$

We conclude that if there is a cycle in $G[N^2(x)\cup R]$ and $x\not\in Q$ then $N(x) \cup Q_2$ is a cct of $G'$, which finishes the proof in Case 1.

\subsection{Case 2: $y$ is a false twin to $x$.}
%{\bf Case 2: $y$ is a false twin to $x$.}

Let $G'$ result from $G$ by adding a false twin $y$ to $x$ in $G$. We again inductively suppose that $G$ has a cct $Q$.
The possible cycles with $y$ in $G'$ are triangles $yab$ for $a,b \in N(x), ab \in E(G)$, $C_4$'s $yabc$ for $a,b \in N(x)$, $ab \notin E(G)$, $c \in N^2(x)$, and $C_4$'s $xyab$ for $a,b \in N(x)$.

If $|N(x)|=1$ then $Q$ is also a cct of $G'$. Now assume that $|N(x)| \ge 2$.

Recall that $V(G)$ is partitioned into $\{x\} \cup N(x) \cup N^2(x) \cup R$, and since $G'$ is $(G_1,G_2,G_3)$-free, $R$ induces a cycle-free subgraph in $G'$.

The fact below strengthens Claim~\ref{N(x)split}.

\begin{claim}\label{N(x)threshold}
$G'[N(x)]$ is a threshold graph.
\end{claim}

{\em Proof of Claim $\ref{N(x)threshold}$.} Since $G'$ is distance hereditary, $N(x)$ is $P_4$-free, and since $G'$ is $G_5$- and $G_6$-free, $N(x)$ is $2K_2$- and $C_4$-free, i.e., $G'[N(x)]$ is a threshold graph which shows Claim~\ref{N(x)threshold}. \hfill $\diamond$

{\bf Case 2.1: $G[N^2(x)\cup R]$ is cycle-free.}

%\begin{claim}\label{ftN2(x)URcyclefree}
%If $y$ is a false twin to $x$ and $G[N^2(x)\cup R]$ is cycle-free then
We are going to show that also in this case, $G'$ has a cct.
%\end{claim}

%{\em Proof of Claim $\ref{ftN2(x)URcyclefree}$.}

Recall that $G$ has a cct $Q$, and let $Q_1=Q \cap N(x)$, $Q_2=Q \cap N^2(x)$, $N_1(x)=N(x)\setminus Q_1$. As in Claim~\ref{ttN2(x)URcyclefree}, $Q_2$ can contain at most two vertices, and we can assume that every vertex in $Q_2$ has a neighbor in $N_1(x)$. Moreover, if $Q_2 \neq \emptyset$ then $x \notin Q$.

{\bf Case 2.1.1: $|Q_2|=2$.}

Let $Q_2=\{u,v\}$; recall that by Proposition \ref{sameneighbors}, $u$ and $v$ have the same neighborhood in $N_1(x)$. We distinguish between three subcases:

$(i)$ If $u,v$ have three neighbors $a,b,c$ in (the stable set) $N_1(x)$ then vertices $x,y,u,a,b,c$ (vertices $x,y,v,a,b,c$, respectively) induce $G_4$ in $G'$, which is impossible.

$(ii)$ If $u,v$ have exactly two neighbors $a,b$ in $N_1(x)$ then there is no other vertex $c \in N_1(x)$, otherwise $x,y,a,b,c,u,v$ induce $G_{11}$ in $G'$. In addition, since $G'$ is $G_4$-free and by Claim~\ref{N(x)threshold}, $G'[N(x)]$ is a threshold graph, either $a$ or $b$ sees all vertices of $Q_1$, otherwise if $a$ misses $a'$ and $b$ misses $b'$ in $Q_1$, respectively, then either $G'$ contains $G_4$ (if we can choose $a'=b'$) or there is a $P_4$ $ab'a'b$ in $N(x)$. Suppose that $a$ sees all vertices of $Q_1$. Then every cycle in $G'$ containing $y$ also contains some vertex in $Q \cup \{a\}$, showing that $Q \cup \{a\}$ is a cct of $G'$.

$(iii)$ If $u,v$ have exactly one neighbor $a$ in $N_1(x)$, we analyze the neighborhood of $a$. If $a$ misses some vertex $b\in Q_1$ then there is no other vertex $c \in N_1(x)$ (otherwise $x,y,a,b,c,u,v$ induce $G_{11}$ if $bc \notin E(G)$ or $c, b, x, u, a$ induce  house if $bc \in E(G)$), and hence every cycle in $G'$ containing $y$ also contains some vertex in $Q$, i.e., $Q$ is still a cct of $G'$. If $a$ sees all vertices in $Q_1$, every cycle containing $u$ or $v$ also contains some vertex in $Q_1\cup\{a\}$, and hence $(Q\setminus\{u,v\})\cup\{a,y\}$ is a cct of $G'$.

{\bf Case 2.1.2: $|Q_2|=1$.}

Let $Q_2=\{u\}$. Since $G'$ is $G_4$-free, $u$ has at most two neighbors in $N_1(x)$. If $u$ has two neighbors $a,b$ in $N_1(x)$ then, by Claim~\ref{N(x)threshold} and since $G'$ is $G_4$-free, one of them, say $a$, must see all vertices in $Q_1$, and this means that every cycle containing $u$ also contains some vertex in $Q_1\cup\{a\}$ (recall that $G[N^2(x)\cup R]$ is cycle-free), i.e., $(Q\setminus\{u\})\cup\{a,y\}$ is a cct of $G'$. If $u$ has precisely one neighbor $a$ in $N_1(x)$ and $a$ sees all vertices in $Q_1$, again $(Q\setminus\{u\})\cup\{a,y\}$ is a cct of $G'$; otherwise, $a$ misses a vertex $b$ in $Q_1$, and the analysis is as follows:

$(i')$ If $N_1(x)$ consists only of vertex $a$ then every cycle in $G'$ containing $y$ also contains a vertex of $Q_1$, and hence $Q$ is a cct of $G'$.

$(ii')$ If $N_1(x)$ contains a vertex $c\neq a$, we must have $bc\not\in E(G)$ (otherwise $x,a,b,c,u$ induce a house). We show that there is no cycle $C$ in $G$ containing $a,u$ but no vertex of $Q_1$. If there is such a cycle $C$ then by Proposition~\ref{sameneighbors}, it must be a triangle $auv$ with $v\in N^2(x)$, and then vertices $x,y,a,b,c,u,v$ induce graph $G_{11}$, or it is a $C_4$ $auvw$ with $u,v \in N^2(x)$ and $w \in R$ but then there is a $G_{12}$ or domino in $G'$. We conclude that $(Q\setminus\{u\})\cup\{y\}$ is a cct of $G'$.

{\bf Case 2.1.3: $Q_2=\emptyset$.}

In this case, $Q\subseteq N[x]$. If there is a cct $Q$ of $G$ with $x\not\in Q$ then $Q \cup \{y\}$ is a cct of $G'$ and we are done. So we have to show that in Case 2.1.3, $G$ has a cct $Q$ with $Q \subseteq N(x)$.
% In this case, Lemma \ref{blocks} plays a fundamental role.

In $G$, there are two types of cycles containing $x$: Triangles $xab$ with $a,b \in N(x)$ and $C_4$'s $xabc$ with $a,b \in N(x)$ and $c \in N^2(x)$. Recall that by Claim \ref{N(x)threshold}, $N(x)$ induces a threshold graph and in particular is partitioned into a clique $Q_1$ and a stable set $N_1(x)$. Then $Q_1$ (and in general, every maximal clique in $N(x)$) covers every triangle $xab$ since $ab \in E$.

The case of $C_4$ with $x$ in $G$ is more involved. Assume that there is a $C_4$ $xabc$ with $a,b \in N(x)$ and $c \in N^2(x)$ which is not covered by $Q_1$. If vertex $a$ (vertex $b$, respectively) sees all vertices in $Q_1$ then the clique $Q_1 \cup \{a\}$ ($Q_1 \cup \{b\}$, respectively) covers $xabc$ as well. Otherwise both $a$ and $b$ have non-neighbors in $Q_1$. Since by Claim \ref{N(x)threshold}, $N(x)$ is $P_4$-free, $a$ and $b$ have a common non-neighbor, say $d$, in $Q_1$. It follows that $cd \notin E$ (otherwise $x,y,c,a,b,d$ induce $G_4$). If $a$ and $b$ miss another vertex $d' \in Q_1$ then $x,y,d,d',a,b,c$ induce $G_{11}$, a contradiction. Thus one of $a$ and $b$, say $a$, has at most one non-neighbor, say $d$, in $Q_1$, and since $N(x)$ is a threshold graph, without loss of generality, the neighborhood of $b$ in $Q_1$ is contained in the neighborhood of $a$ in $Q_1$; in particular, $b$ misses $d$, and, as above, $c$ misses $d$. Let $e$ be a neighbor of $d$ in $N^2(x)$.

\begin{claim}\label{misses}
Let $e$ be a neighbor of $d$ in $N^2(x)$. Then $e$ misses $a, b$ and $c$. (with $a$, $b$, $c$ and $d$ as described above).
\end{claim}

{\em Proof of Claim $\ref{misses}$.}

We begin by observing that if  $ce \in E$, by Proposition 1 (i), $c$ and $e$ have the same neighbors in $N(x)$ which is impossible since $e$ sees $d$ and $c$ misses $d$. Therefore $ce \not\in E(G)$. In this case, by Proposition 1 (ii), if $e$ sees one of $a$ and $b$, it must see both of them but now, $x,y,e,a,b,d$ induce $G_4$ - a contradiciton. Then $e$ must also miss both $a$ and $b$.
 \hfill $\diamond$

Suppose that $Q'_1:=(Q_1 \setminus \{d\}) \cup \{a\}$ is not a cct of $G$. Note that $N(x) \setminus Q'_1$ is stable. Then there is a cycle in $G$ whose only vertex from $Q_1$ is $d$. Obviously, if $C$ is a cycle containing $d$ and an edge in $N(x)$ then $Q'_1$ covers $C$ since $N(x) \setminus Q'_1$ is stable. Thus, we have to consider cycles without an edge in $N(x)$.

First consider a $C_3$ $duv$ with $u,v \in N^2(x)$. Then by Claim \ref{misses}, $u$ and $v$ miss $a,b$ and $c$, and now, together with $y$, $G'$ contains $G_9$, a contradiction. If $d$ is in a $C_4$ $duvw$ with $u,v \in N^2(x)$ and $w \in R$ then very similarly, together with $y$, $G'$ contains $G_{10}$, a contradiction. Thus, $d$ is not contained in any of such cycles.

If $C$ is a $C_4$ with $d,z \in N(x)$ and $u,v \in N^2(x)$ then, again by Claim \ref{misses}, $u$ and $v$ miss $a,b$ and $c$. Then $z$ must see $a$ and $b$, otherwise there is a house or $G_{12}$ in $G'$, together with $y$, but then $xadzu$ induce a house, a contradiction.

This also happens when $d$ is in a $C_4$ with $x$ and no vertex from $Q'_1$. This final contradiction shows that in Case 2.1.3, there is a cct $Q$ of $G$ without $x$, and thus, there is a cct $Q \cup \{y\}$ in $G'$.

{\bf Case 2.2: $G[N^2(x)\cup R]$ is not cycle-free.}

As in Case 1, we now assume that $G[N^2(x)\cup R]$ contains a cycle $C$ (which implies that in this case, $x\not\in Q$ holds).

\begin{claim}\label{alphaN(x)le2}
%If $y$ is a false twin to $x$ and $G[N^2(x) \cup R]$ is not cycle-free then
$N(x)$ is $I_3$-free.
\end{claim}

{\em Proof of Claim $\ref{alphaN(x)le2}$.}
Assume that $G[N^2(x) \cup R]$ is not cycle-free and $N(x)$ contains a stable set of three vertices $a_1,a_2,a_3$. Let $S$ be a maximal stable set in $N(x)$ containing $a_1,a_2,a_3$. Recall that by Proposition~\ref{sameneighbors}, for vertices $u,v \in N^2(x)$ in the same connected component of $G[N^2(x) \cup R]$, their neighborhoods in $S$ are equal. In addition, no vertex $u \in N^2(x)$ sees at least three vertices in $S$, otherwise $G_4$ is contained in $G'$. Thus, for every pair of vertices $u,v \in N^2(x)$, $N(u) \cap S = N(v) \cap S \subseteq \{a_1,a_2\}$ holds.

If $u \in N^2(x)$ is in a cycle of type ($A_1$) or ($A_2$) then $x,y,a_1,a_2,a_3,u$ and the vertices of the remaining cycle induce $G_2, G_3, G_9$ or $G_{10}$. If the cycle with $u,v \in N^2(x)$ is of type ($B_1$), there is a house; if it is of type ($B_2$) or ($D_1$), there is $G_2$ or $G_{11}$; if of type ($B_3$), there is $G_3$ or $G_{12}$; and finally, if of type ($C_1$) or ($D_2$), there is $G_3$ or $G_{11}$. This shows Claim~\ref{alphaN(x)le2}. \hfill $\diamond$

We conclude that if $y$ is a false twin to $x$ and $G[N^2(x) \cup R]$ is not cycle-free then by Claim $\ref{alphaN(x)le2}$, $N_1(x)$ contains at most two vertices. If $|N_1(x)|\leq 1$ then $Q$ is a cct of $G'$. If $N_1(x)=\{a,b\}$ then by Claims $\ref{N(x)threshold}$ and \ref{alphaN(x)le2},
one of them, say $a$, sees all vertices in $Q_1$. By Proposition~\ref{sameneighbors}, either $Q_2\cup\{a\}$ is a clique, and then $Q\cup\{a\}$ is a cct of $G'$,
or $a$ sees no vertex of $Q_2$, and then let $C$ be a cycle in $G[N^2(x) \cup R]$ and let $u \in V(C)\cap Q_2$. Since $xyab$ is a $C_4$ and $G$ is $(G_2,G_3)$-free, there is an edge linking $xyab$ and $C$. By Proposition~\ref{sameneighbors}, we conclude that $b$ sees all vertices in $C \cap N^2(x)$ and, therefore, in $Q_2$. Now if there is some $a' \in Q_1$ then $x,a,a',b,u$ induce either a house or a gem -- a contradiction. Therefore $Q_1=\emptyset$ and $Q'_2 \cup\{b\}$ is a cct of $G'$. This finishes the proof in Case 2 and thus also the proof of Theorem~\ref{maintheocctdhg}. \qed

As a direct consequence of Theorem~\ref{maintheocctdhg}, we obtain another proof of a result in~\cite{BraBriKleNogPro2011}:

\begin{corollary}\label{forb-cographs}
If $G$ is a cograph then $G$ admits a clique cycle-transversal if and only if $G$ is $(G_1,\ldots,G_6)$-free.
\end{corollary}

\proof Graphs $G_1$ to $G_6$ admit no cct. Conversely, $G$ is also $(G_7,\ldots,G_{12})$-free
(because all of them contain $P_4$). Since every cograph is a distance-hereditary graph,
by Theorem~\ref{maintheocctdhg} the corollary follows. \qed

\begin{corollary}\label{forb-(2,1)-DHG}
Let $G$ be a distance-hereditary graph. Then $G$ is a $(2,1)$-graph if and only if $G$ is $(G_1, G_5, G_6, G_7)$-free.
\end{corollary}

\proof Graphs $G_1, G_5, G_6$ and $G_7$ are not $(2,1)$-graphs. Conversely, assume that $G$ is $(G_1, G_5, G_6, G_7)$-free and $G$ is not a $(2,1)$-graph. Let $G'$ be a minimal induced subgraph of $G$ which is not a $(2,1)$-graph. Note that being a $(2,1)$-graph is equivalent to admitting a clique that intersects every odd cycle. Thus $G'$ does not admit a cct. By Theorem~\ref{maintheocctdhg}, $G'$ is isomorphic to one of the graphs $G_1, G_2, \ldots, G_{12}$. Since $G_2, G_3, G_4, G_8, G_9, G_{10}, G_{11}$, and $G_{12}$ are $(2,1)$-graphs, it follows that $G$ contains $G_1, G_5, G_6$, or $G_7$ as an induced subgraph. \qed

\if 10

Similarly, one can recognize in linear time whether a given distance-hereditary graph has a stable cycle transversal.

Characterizing distance-hereditary graphs having an sct in terms of forbidden subgraphs or in another way remains an open question. We remark that there are infinite families of
vertex-minimal distance-hereditary graphs with no sct. Some examples are shown in
Figure~\ref{fig:forb-sct-dhg}. The graph on the left is not an $(\S,\F)$-graph because no sct
can contain $a,b$ or $c$. On the right, no sct can contain $d,e,f$ or $g$ and thus there are $C_4$'s left.

\begin{figure}[tb]
\begin{center}
\includegraphics[width=13cm]{forb-sct-dhg.pdf}
\caption{Vertex-minimal distance-hereditary graphs with no sct.}\label{fig:forb-sct-dhg}
\end{center}
\end{figure}

\fi

\end{document}